\documentclass[twocolumn]{article}

\usepackage[english]{babel}

\usepackage{geometry}
\usepackage{amsmath}
\usepackage{graphicx}
\usepackage[colorlinks=true, allcolors=blue]{hyperref}
\usepackage[OT1]{fontenc} 
\usepackage{balance}
\usepackage{anyfontsize}
\usepackage{microtype}
\usepackage{graphicx}
\usepackage{booktabs} 
\usepackage{hyperref}

\usepackage{tablefootnote}
\usepackage{amsmath}
\usepackage{mathtools}
\usepackage{amsthm}
\usepackage{booktabs}
\usepackage{amsfonts}
\usepackage{amsmath}
\usepackage{multirow}
\usepackage{amsthm}
\usepackage{diagbox}
\usepackage{subcaption}
\usepackage{bm}
\usepackage{color}
\usepackage{graphicx}
\usepackage{natbib}
\usepackage{enumitem}
\usepackage{algorithm}
\usepackage{algorithmic}
\usepackage{mathtools}
\usepackage{tablefootnote}
\usepackage{thmtools} 
\usepackage{thm-restate}
\usepackage{xcolor}         
\definecolor{pku-red}{RGB}{139,0,18}
\usepackage[capitalize,noabbrev,nameinlink]{cleveref}
\usepackage[textsize=tiny]{todonotes}
\usepackage{bbm}
\usepackage[normalem]{ulem}

\theoremstyle{plain}

\theoremstyle{definition}

\theoremstyle{remark}

\newcommand{\EE}{\mathbb{E}}

\newcommand{\RR}{\mathbb{R}}

\newcommand{\bmC}{{\bm{C}}}

\newcommand{\bmR}{{\bm{R}}}

\newcommand{\bma}{{\bm{a}}}

\newcommand{\bmh}{{\bm{h}}}

\newcommand{\bmp}{{\bm{p}}}

\newcommand{\bms}{{\bm{s}}}

\newcommand{\bmv}{{\bm{v}}}

\newcommand{\bmx}{{\bm{x}}}

\newcommand{\CPR}{{\mathrm{CPR}}}
\newcommand{\GRU}{{\mathrm{GRU}}}
\newcommand{\MLP}{{\mathrm{MLP}}}

\title{\bfseries{An Adaptable Budget Planner for Enhancing Budget-Constrained Auto-Bidding in Online Advertising}}

\newcommand{\email}[1]{\texttt{#1}}

\author{
	\textbf{Zhijian Duan}$^{1,2,*}$, 
	\textbf{Yusen Huo}$^{3}$,
	\textbf{Tianyu Wang}$^{3}$,
	\textbf{Zhilin Zhang}$^{3}$,
	\textbf{Yeshu Li}$^{3}$, \\
	\textbf{Chuan Yu}$^{3}$,
	\textbf{Jian Xu}$^{3}$,
	\textbf{Bo Zheng}$^{3}$,
	\textbf{Xiaotie Deng}$^{1,2}$ 
	\\
	$^{1}$School of Computer Science, Peking University, Beijing, China \\
	$^{2}$Center on Frontiers of Computing Studies, Peking University, Beijing, China \\
	$^{3}$Alibaba Group, Beijing, China
	\\
	\email{zjduan@pku.edu.cn}, 
	\\
	\email{\{huoyusen.huoyusen, yves.wty, zhangzhilin.pt\}@alibaba-inc.com}, 
	\\
	\email{\{liyeshu.lys, yuchuan.yc, xiyu.xj, bozheng\}@alibaba-inc.com},
	\\
	\email{xiaotie@pku.edu.cn}
}

\date{}

\begin{document}
	\maketitle
	\def\thefootnote{*}\footnotetext{This work is done during internship at Alibaba Group.}

	\begin{abstract}
In online advertising, advertisers commonly utilize auto-bidding services to bid for impression opportunities. A typical objective of the auto-bidder is to optimize the advertiser's cumulative value of winning impressions within specified budget constraints. However, such a problem is challenging due to the complex bidding environment faced by diverse advertisers. To address this challenge, we introduce ABPlanner, a few-shot adaptable budget planner designed to improve budget-constrained auto-bidding. ABPlanner is based on a hierarchical bidding framework that decomposes the bidding process into shorter, manageable stages. Within this framework, ABPlanner allocates the budget across all stages, allowing a low-level auto-bidder to bids based on the budget allocation plan. The adaptability of ABPlanner is achieved through a sequential decision-making approach, inspired by in-context reinforcement learning. For each advertiser, ABPlanner adjusts the budget allocation plan episode by episode, using data from previous episodes as prompt for current decisions. This enables ABPlanner to quickly adapt to different advertisers with few-shot data, providing a sample-efficient solution. Extensive simulation experiments and real-world A/B testing validate the effectiveness of ABPlanner, demonstrating its capability to enhance the cumulative value achieved by auto-bidders.
\end{abstract}

	\section{Introduction}
Real-Time Bidding (RTB) plays a crucial role in online advertising, where an ad auction is triggered in real-time whenever an ad impression opportunity arises~\citep{ou2023survey}. 
The speed and efficiency of RTB enable advertisers to target specific audiences and optimize their campaigns in real time. 
To navigate the complex and dynamic nature of the advertising landscape, advertisers often turn to auto-bidding strategies~\citep{aggarwal2019autobidding}. 
Auto-bidding systems leverage algorithms and machine learning models to make rapid and informed bidding decisions on behalf of advertisers. 

One primary objective of auto-bidders is to maximize the cumulative value of winning impressions for advertisers while adhering to specified budget constraints~\citep{balseiro2019learning, chen2023coordinated}. 
Since the values and prices of arriving impressions are initially unknown, such a budget-constrained bidding can be viewed as an online stochastic knapsack problem~\citep{hao2020dynamic}. 
In this context, auto-bidders ideally aim to win impressions with high values and low market prices, that is, impressions with high \emph{cost-performance ratios}, to effectively achieve their objectives~\citep{zhou2008budget, lin2016combining, maehara2018optimal}.
However, this task is challenging due to the significant randomness in the fine-grained information of each incoming impression, and the bidding process within an episode involves numerous impressions. 
Additionally, different advertisers faces different bidding environments, which complicates the optimization of bidding strategies for each individual advertiser~\citep{zhang2023personalized}.

To address the challenges above, in this paper, we introduce ABPlanner, a few-shot \underline{A}daptable \underline{B}udget \underline{Planner} to enhance budget-constrained auto-bidding in online advertising.
ABPlanner is based on a hierarchical bidding framework that decomposes the long bidding episode into short stages, each representing a fragment within the overarching process.
In the hierarchical bidding framework, the high-level budget planner allocates the budget across these stages, and the underlying auto-bidder makes decisions within each stage based on the budget allocation plan.
Such a framework enables us to capture the temporal variation of impressions through the high-level budget planner, which can strategically allocate more budget to stages with higher expected cost-performance ratios and less budget to stages with lower expected ratios. 
This strategy is feasible since, compared to a single impression, the cross-grain information of impressions within a stage exhibits less randomness. 
Additionally, we mitigate decision-making complexities for the underlying auto-bidder, as it only needs to make decisions within each short stage rather than across the entire lengthy bidding episode.

Based on the hierarchical bidding framework, ABPlanner is designed as a sequential decision-maker to achieve adaptability, inspired by in-context reinforcement learning~\citep{laskin2023incontext, lee2024supervised}. 
Specifically, we model ABPlanner's decision-making process as a Markov Decision Process (MDP). 
In this MDP framework, each bidding episode represents a decision step, and each adjustment of the allocation plan is treated as an action.
As a result, ABPlanner can adjust its budget allocation plan for each advertiser episode by episode, utilizing data from previous episodes as prompts~\citep{xu2022prompting} for current decisions. 
This design enables ABPlanner to rapidly adapt to new advertisers, making it a sample-efficient solution.
Within this MDP framework, ABPlanner can be trained using standard deep reinforcement learning algorithms, such as Proximal Policy Optimization (PPO)~\citep{schulman2017proximal}.

To assess the effectiveness and adaptability of our approach, we conduct comprehensive experiments, comprising simulation and real-world A/B testing. 
The simulation experiments include a pure simulation environment based on synthetic data and a semi-simulation environment utilizing real-world data.
Both the experimental results demonstrate the advantage of introducing a high-level budget planner to the underlying auto-bidder and the benefits of ABPlanner to adapt to previously unseen advertisers. 
Moreover, in real-world A/B testing, we deploy ABPlanner to a real-world advertising system. 
The results indicate that ABPlanner significantly improves the cumulative value of the underlying auto-bidder.

In summary, our main contributions are:
\begin{itemize}
    \item We employ a hierarchical bidding framework that augments the auto-bidder with a high-level budget planner. This planner strategically allocates the budget across decomposed stages of the bidding episode, effectively capturing temporal variations in impressions.

    \item We propose ABPlanner, a few-shot adaptable budget planner modeled as a sequential decision-maker within a Markov Decision Process (MDP). ABPlanner dynamically adjusts budget allocation based on prompts derived from few-shot episode data generated through online interactions, thereby enhancing sample efficiency and applicability to new advertisers.

    \item We validate the effectiveness and adaptability of our proposed approach through extensive experiments. Deployment and justification in a real-world advertising system further confirm the practical utility of ABPlanner.
\end{itemize}

Our paper is structured as follows: 
In \cref{sec:RelatedWork}, we review the related works. 
Next, we delve into the budget-constrained auto-bidding problem and hierarchical bidding framework in \cref{sec:HierarchicalBiddingFramework}. 
Following that, \cref{sec:AdaptableBudgetPlanner} provides a detailed explanation of ABPlanner and its learning algorithm. 
We present the results of our experiments in \cref{sec:Experiment}, and finally, we conclude our paper in \cref{sec:Conclusion}.

\section{Related Work}
\label{sec:RelatedWork}

\paragraph{Bid Optimization}
Bid optimization in online advertising is a sequential decision procedure~\citep{xu2024auto, he2021unified, su2024auctionnet}. 
Control-based methods, such as \citet{zhang2016feedback}, \citet{yang2019bid} and traditional Proportional-Integral-Differential (PID) controllers~\citep{bennett1993development}, utilize online feedback to regulate the bidding strategy. 
The auto-bidding problem can also be formulated as a Markov Decision Process (MDP) and solved using reinforcement learning (RL). 
On top of that, studies by \citet{wu2018budget}, \citet{zhao2018deep}, \citet{he2021unified}, \citet{guan2021multi} and \citet{yuan2022actor} employ RL to optimize bidding policies under various constraints. 
Furthermore, \citet{mou2022sustainable} design an iterative offline RL framework to address the sim2real problem, and \citet{guo2024generative} introduce a conditional diffusion model to solve the MDP-based bidding problems.
Despite their efficacy in optimizing bidding strategies, these methods typically implicitly capture the temporal distribution of impressions.

\paragraph{Bid with Prediction}
Since the challenge of auto-bidding comes from the uncertainty of arriving impressions, several approaches leverage auxiliary prediction tasks. 
A common task involves predicting the winning price of an arrived impression. 
For instance, \citet{lin2016combining} introduce an additional winning price predictor to optimize real-time bidding strategy. 
Similarly, \citet{zhou2021efficient} utilize a deep distribution network to model winning price distribution explicitly. 
Additionally, \citet{ou2023deep} address multi-slot real-time bidding by predicting both the winning price distribution and the winning probability of bid prices for all ad positions.
Another approach involves modeling the bidding environment of the auto-bidder. 
\citet{cai2017real} propose a model-based RL algorithm to optimize bidding policy by modeling the state transition in each arriving ad auction. 
Extending this idea, \citet{chen2023model} consider relative coarse-grained state transitions, corresponding to periods encompassing numerous impressions. 
However, these auxiliary prediction tasks remain challenging due to the substantial randomness of impressions within a bidding episode.

\paragraph{Multi-Channel Bidding}
\label{subsec:MultiChannelBidding}
A similar hierarchical framework, featuring a high-level planner and a low-level auto-bidder, is prevalent in the domain of multi-channel bidding~\citep{deng2023multi}, also known as multi-platform bidding~\citep{avadhanula2021stochastic} or campaign optimization~\citep{zhang2012joint}. 
In this scenario, advertisers allocate their budget across various channels, and the auto-bidder in each channel concurrently bids under the specified budget. 
Empirically, \citet{li2018efficient} and \citet{xiao2019model} employ reinforcement learning to derive budget allocation strategies for the planner. 
\citet{luzon2022dynamic} develop closed-form solutions for dynamic budget allocation based on a fitted effectiveness function. 
Additionally, \citet{nuara2022online} and \citet{wang2023hibid} jointly learn high-level budget allocation and low-level bidding strategies through offline reinforcement learning.
Compared to these empirical works that solve the budget allocation plan for a single upcoming episode, our proposed ABPlanner adaptively adjusts the budget allocation plan across successive episodes.

\section{Hierarchical Bidding Framework}
\label{sec:HierarchicalBiddingFramework}

In this section, we introduce the budget-constrained bidding problem in online advertising and the hierarchical bidding framework that enhances an auto-bidder with a high-level budget planner. 

\subsection{Budget-Constrained Auto-Bidding}

We consider the problem of budget-constrained auto-bidding, where an auto-bidder participates in sequential auctions on behalf of an advertiser to maximize the cumulative value of winning impressions under a specified budget constraint.
Let $n$ represent the (typically unknown) number of sequentially arriving impression opportunities, with $i \in [n]$ indexing these opportunities. 
The auto-bidder aims to secure impressions that maximize cumulative value under the maximum budget limitation of $B$. 

At each arriving impression $i \in [n]$, the auto-bidder observes the value $v_i$ of the impression and submits a bid $b_i$. 
It wins the impression only if its bid $b_i$ exceeds other advertisers' highest bid $p_i$. 
Upon winning impression $i$, the advertiser receives the impression's value $v_i$ and pays the price $p_i$. 
We denote by $x_i \coloneqq \mathbbm{1}\{b_i > p_i\}$ the binary indicator of whether the auto-bidder wins impression $i$.

Formally, the process can be represented as follows:
\begin{equation}\label{eq:OPT:bidding}
\begin{aligned}
    \max_{\bmx \in \{0, 1\}^n}~&\mathbb{E}_{\bmv,\bmp}\left[\sum_{i=1}^n x_i v_i\right], &&
    \text{s.t.}~\sum_{i=1}^n x_i p_i \le B,
\end{aligned}
\end{equation}
where the expectation is taken with respect to the random values $\bmv = (v_1, v_2, \dots, v_n)$ and prices $\bmp = (p_1, p_2, \dots, p_n)$ of the arriving impressions.

The key to solving \cref{eq:OPT:bidding} is to win impressions with high values and low payment prices, that is, impressions with high cost-performance ratios~\citep{hao2020dynamic}, which defined as $\CPR_i \coloneqq \frac{v_i}{p_i}$ for impression $i$.
However, the values and prices of arriving impressions are initially unknown, making \cref{eq:OPT:bidding} an online stochastic knapsack problem.
Additionally, upon the arrival of an impression $i$, the auto-bidder can only observe its value $v_i$ before making a decision, and the price $p_i$ is revealed if the auto-bidder wins impression $i$. 
Therefore, it is crucial for the auto-bidder to identify impressions with high cost-performance ratios in advance.

Capturing the temporal distribution of cost-performance ratios poses a challenge for the auto-bidder due to two reasons.
Firstly, the value and price distributions of arriving impressions exhibit significant randomness, making it challenging to predict the cost-performance ratio. 
Secondly, the bidding process within an episode involves a large number of impressions, transforming it into a long-horizon sequential decision-making problem.

\subsection{Hierarchical Bidding Framework}


To alleviate the randomness and long-horizon issues, we reformulate the budget-constrained auto-bidding problem as a hierarchical framework, which enhances the auto-bidder with a \emph{high-level budget planner}.
We initially partition the bidding episode into $m$ stages. 
This division can be based on equal time intervals or the quantity of impressions.
Additionally, we incorporate a high-level budget planner to establish the budget plan $\bm\rho = (\rho_1, \rho_2, \dots, \rho_m)$ for all these stages. 
Each planned budget $\rho_i$ can serve as either the allocated budget for the $i$-th bidding stage or as reference budget consumption, acting as side information for the auto-bidder.
Subsequently, the underlying auto-bidder places bids under the budget plan $\bm\rho$.

Under this hierarchical framework, the planner establishes a budget plan within the budget constraint for an advertiser $ c$.
The planner's goal is to maximize the expected cumulative values obtained by the auto-bidder. 
The problem can be formulated as:
\begin{equation}
    \label{eq:OPT}\tag{OPT}
    \begin{aligned}
        \max_{\bm\rho}~& \sum_{i=1}^m R_{c,i}(\bm\rho), &&
        \text{s.t.}~\sum_{i=1}^m \rho_i \le B,
    \end{aligned}
\end{equation}
where $R_{c,i}(\bm\rho)$ represents the expected return obtained by the auto-bidder in the $i$-th stage, given the entire budget plan $\bm\rho$.
Note that we provide $\bm\rho$ instead of $\rho_i$ to the auto-bidder in stage $i$ as the auto-bidder can use the complete budget plan as side information to make decisions. 
Additionally, the expectation considers any randomness in the auto-bidder and the random values of the advertiser.

Note that we \emph{do not restrict} the specific bidding strategy of the low-level auto-bidder.
It can be any bidding algorithm that operates under the budget constraint. 
In our paper, our emphasis is on enhancing the auto-bidder through the learning of the high-level budget planner.

Introducing a budget planner offers three key advantages in budget-constrained auto-bidding. 
Firstly, compared to each single impression, the cross-grain statistical information of impressions within a stage exhibits less randomness. 
Thus, the budget planner's decision environment is more stable and less complex than that of the auto-bidder, making it easier to learn. 
Secondly, this relatively stable decision environment makes it easier for the budget planner to capture the temporal distribution of cost-performance ratios. 
Specifically, the budget planner can strategically allocate more budget to stages with higher expected cost-performance ratios and less budget to stages with lower expected ratios. 
Thirdly, the decision-making complexities for the underlying auto-bidder are mitigated, as decisions only need to be made within each short stage under the budget plan rather than across the entire long-horizon bidding episode. 
This reduction in decision-making complexity can lead to a more efficient and effective auto-bidder.

\section{Adaptable Budget Planner}
\label{sec:AdaptableBudgetPlanner}

\begin{figure*}[t]
    \centering
    \includegraphics[width=\textwidth]{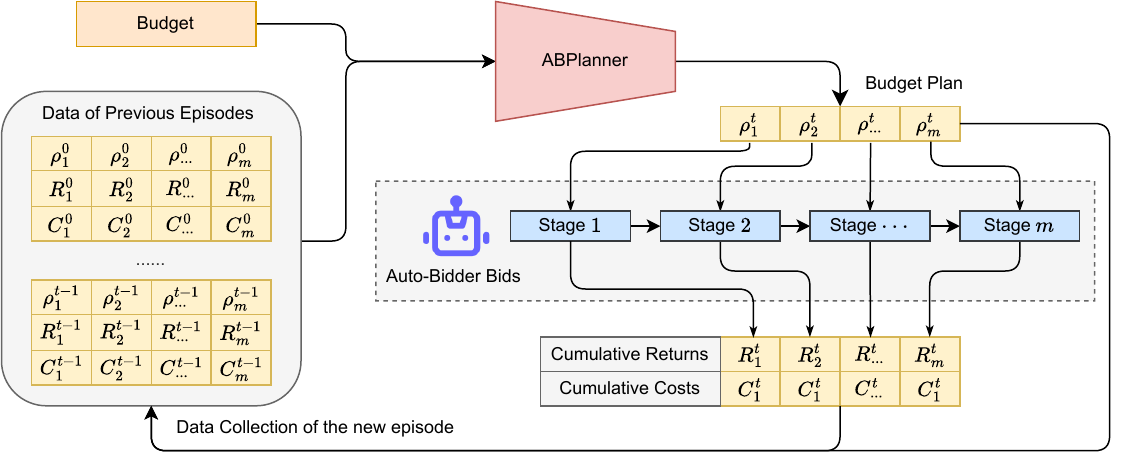}
    \caption{
        The whole procedure of ABPlanner. 
        At the beginning of each bidding episode of an advertiser, ABPlanner allocates the budget based on  data from previous bidding episodes.
        Subsequently, the underlying auto-bidder bids according to the allocation plan, and ABPlanner collects information to inform future episodes for the advertiser.
    }
    \label{fig:ABPlanner}
\end{figure*}
A straightforward approach to allocate the budget is to fit the budget-return function $R_{c,i}(\rho_i)$ for all the stages and solve an optimization problem to determine the budget allocation plan.
However, this method encounters a sample efficiency challenge, as it requires substantial historical data from advertisers to fit their own budget-return function.
This issue hinders the adaptability of the budget planner to novel scenarios or diverse advertisers with limited historical data.

To make a sample-efficient budget planner, we propose ABPlanner, a few-shot \emph{adaptable} budget planner designed to enhance budget-constrained auto-bidding. 
ABPlanner is modeled as a sequential decision-maker within a Markov Decision Process (MDP), dynamically adjusting an advertiser's budget plan across successive episodes. 
By utilizing data from previous episodes as prompts for current decisions, ABPlanner addresses challenges related to insufficient or outdated bidding logs. 
The entire process of ABPlanner is summarized in \cref{fig:ABPlanner}.

\subsection{Markov Decision Process Modeling}

Given an advertiser indexed by $c$, we first receive the initial budget allocation plan $\bm\rho^0$ at the beginning episode $0$. 
$\bm\rho^0$ can be initialized through equal division or based on the historical budget consumption of each stage. 
Subsequently, ABPlanner sequentially adjusts the daily budget plan, modeling this adjustment process as a Markov Decision Process (MDP). 
The key components are:
\begin{itemize}
    \item \textbf{State} $\bms_t = (B, \bm\rho^{0:t-1}, \bmR_c^{0:t-1}, \bmC_c^{0:t-1})$, which includes the budget $B$, the budget plan $\bm\rho^{0:t-1}$, the cumulative returns $\bmR_c^{0:t-1}$, and the cumulative costs (payments) $\bmC_c^{0:t-1}$ for all stages in the preceding episode $0$ to $t-1$ of advertiser $c$.
    
    \item \textbf{Action} $\bma_t \in \mathbb{R}^m$: 
    The action $\bma_t$ represents the update direction of the budget allocation plan.
    Utilizing action $\bma_t$ and the preceding budget plan $\bm\rho^{t-1}$, we update the budget plan by projecting the new allocation onto the simplex $\Delta_B \coloneqq \{ \bm\rho \in \mathbb{R}_{\ge 0}^m: \sum_{i=1}^m \rho_i \le B \}$.
    Formally, 
    \begin{equation*}
        \bm\rho^t = \Pi_{\Delta_B}(\bm\rho^{t-1} + \bma_t) \in \Delta_B,
    \end{equation*}
    where $\Pi_{\Delta_B}$ is the projection operator onto the simplex $\Delta_B$.

    \item \textbf{Transition}: 
    Given the budget allocation plan $\bm\rho^t$, the underlying auto-bidder bids in each stage.
    Afterward, we collect the cumulative values $\bmR_c^t$ and costs $\bmC_c^t$ for all $m$ stages.
    
    \item \textbf{Reward} $r_t \in \mathbb{R}$: 
    The reward is defined as the increment in cumulative value compared to the previous episode: 
    \begin{equation*}
        r_t = \sum_{i=1}^m R_{c,i}^t - \sum_{i=1}^m R_{c,i}^{t-1}.
    \end{equation*}
    This formulation encourages the planner to seek improvement in each episode.

\end{itemize}
Given the MDP modeling, ABPlanner aims to maximize the expected cumulative rewards $\EE_{c}~\left[ \sum_{t=1}^T r_t \right]$ over $T$ bidding episodes, where the expectation is taken with respect to randomly sampled advertisers.
Through this MDP modeling, ABPlanner becomes adaptable, quickly adjusting to unseen advertisers and modifying the budget allocation plan based on few-shot interacted data.

\subsection{Learning the Adaptable Budget Planner} 

In addition to the MDP modeling, we present the method for learning ABPlanner using Proximal Policy Optimization (PPO)~\citep{schulman2017proximal}. 
PPO is rooted in the actor-critic framework, where it learns a parameterized policy $\pi_{\bm{\bm\theta}}$ and a value function $V_{\bm\phi}$. 
This on-policy reinforcement learning algorithm alternates between collecting trajectories and updating the parameters.

\subsubsection{Architecture of Policy Network}
The parametrized policy $\pi_{\bm{\bm\theta}}$, or the policy network, maps states into actions. 
Firstly, for state $\bms_t = (B, \bm\rho^{0:t-1}, \bmR_c^{0:t-1}, \bmC_c^{0:t-1})$, we utilize a Multi-Layer Perceptron (MLP) to encode the state and a Gated Recurrent Unit (GRU) network~\citep{cho2014learning} to integrate the representations of all historical budget plans, cumulative returns, and costs:
\begin{equation}\label{eq:GRU}
\begin{aligned}
    \bmh_1 &= \GRU(\bm{0}, \MLP_e(\bm\rho^{0}, \bmR_c^{0}, \bmC_c^{0})) \\
    \bmh_t &= \GRU(\bmh_{t-1}, \MLP_e(\bm\rho^{t-1}, \bmR_c^{t-1}, \bmC_c^{t-1})), \quad t > 1, 
\end{aligned}
\end{equation}
where $\bm{0}$ is a zero vector, and $\bmh_t$ is the low-dimensional embedding of historical information in the $t$-th episode. 
Secondly, based on $\bmh_t$ and budget $B$, we calculate the probability distribution of the action. 
This distribution is modeled as a multivariate Normal distribution with mean vector $\bm\mu \in \RR^{m}$ and covariance matrix $\sigma I \in \RR^{m\times m}$ with $\sigma > 0$.
The mean vector is computed by another MLP:
\begin{equation*}
    \bm\mu = \MLP_m(\bmh_t, B) \in \RR^m,
\end{equation*}
and we set the logarithm of standard deviation $\sigma$ as a learnable parameter. 
Afterward, the action is drawn from $N(\bm\mu, \sigma I)$, i.e., 
\begin{equation*}
    \bma_t = \pi_{\bm{\bm\theta}}(\bms_t) \sim N(\bm\mu, \sigma I) \in \RR^m.
\end{equation*}
In summary, the parameters ${\bm{\bm\theta}}$ of the policy network include all the parameters in $\MLP_e$, $\GRU$, $\MLP_m$, and the logarithm of $\sigma$.

\subsubsection{Architecture of Value Network}
The parameterized value function $V_{\bm\phi}$, or the value network, predicts the discounted cumulative sum of rewards $\sum_{i=t}^{T} \gamma^{i - t} r_t$ given the state $\bms_t$. 
We share the historical information encoder of $\pi_{\bm{\bm\theta}}$ and $V_{\bm\phi}$, reusing the historical embedding $\bmh_t$ from \cref{eq:GRU}. 
Based on $\bmh_t$ and budget $B$, we compute the predicted value through another MLP:
\begin{equation*}
    V_{\bm\phi}(\bms_t) = \MLP_v(\bmh_t, B) \in \RR.
\end{equation*} 
In summary, the parameters ${\bm\phi}$ include all the parameters in $\MLP_e$, $\GRU$, and $\MLP_v$. 

\subsubsection{Training and Inference}
\begin{algorithm}[t]
    \caption{Trajectory Collection of ABPlanner}
    \label{alg:Trajectory}
    \begin{algorithmic}[1]
        \STATE \textbf{Input:} Policy of the high-level budget planner $\pi$, size of trajectory dataset $n_t$.
        \STATE \textbf{Output:} Trajectory dataset $\mathcal{D} = \{\bm{\tau}_i\}_{i=1}^{n_t}$
        \STATE Initialize $\mathcal{D} \gets \emptyset$
        \WHILE{$|\mathcal{D}| < n_t$}
            \STATE Randomly sample an advertiser $c$.
            \STATE Set the initial budget allocation plan $\bm{\rho}_0$.
            \STATE Receive $\bm{R}_c^{0}$ and $\bm{C}_c^{0}$ from the underlying auto-bidder.
            \FOR{$t=1$ \textbf{to} $T$}
                \STATE $\bm{s}_t \gets (B, \bm{\rho}^{0:t-1}, \bm{R}_c^{0:t-1}, \bm{C}_c^{0:t-1})$
                \STATE $\bm{a}_t \gets \pi(\bm{s}_t)$
                \STATE $\bm{\rho}^t \gets \Pi_{\Delta_B}(\bm{\rho}^{t-1} + \bm{a}_t)$
                \STATE Receive $\bm{R}_c^{t}$ and $\bm{C}_c^{t}$ from the underlying auto-bidder.
                \STATE $r_t \gets \sum_{i=1}^m R_{c,i}^t - \sum_{i=1}^m R_{c,i}^{t-1}$
            \ENDFOR
            \STATE $\mathcal{D} \gets \mathcal{D} \cup \{\bm{\tau} \coloneqq (\bm{s}_t, \bm{a}_t, r_t, \bm{s}_{t+1})_{t=1}^T\}$
        \ENDWHILE
        \STATE \textbf{return} $\mathcal{D}$
    \end{algorithmic}
\end{algorithm}
During training, we utilize the well-established Proximal Policy Optimization with Clipping (PPO-clip) algorithm to update the policy parameters ${\bm{\theta}}$ and value parameters ${\bm{\phi}}$. In each iteration of the parameter updating process, the PPO algorithm collects multiple trajectories based on the current policy. The trajectory collection procedure is detailed in \cref{alg:Trajectory}. Using the newly gathered trajectory dataset, the PPO algorithm updates the policy parameters ${\bm{\theta}}$ through importance sampling and adjusts the value parameters ${\bm{\phi}}$ via the temporal difference loss.

During inference, ABPlanner operates significantly faster than the low-level auto-bidder does. This is because ABPlanner only makes decisions at the beginning of each upcoming bidding episode, whereas the low-level auto-bidder makes decisions at every step of the bidding. Since ABPlanner only utilizes GRU and MLP, it requires $O(mH)$ computations in each decision-making step, where $m$ is the number of stages, and $H$ is the number of hidden nodes in MLP as well as GRU.

Notably, ABPlanner does not impose restrictions on the bidding strategy of the underlying auto-bidder or the bidding order of all stages. 
Therefore, ABPlanner can also be applied in multi-channel bidding scenarios (discussed in \cref{subsec:MultiChannelBidding}), where advertisers need to allocate their budgets among multiple channels, and the auto-bidder in each channel can bid concurrently under the specified budget allocation.

\section{Experiment}
\label{sec:Experiment}

To evaluate the effectiveness and adaptability of ABPlanner\footnote{Our implementation is available at \url{https://github.com/zjduan/ABPlanner}.}, we conducted extensive experiments, including both simulation experiments and real-world A/B testing.

\subsection{Simulation Experiment}

In the simulation experiments, we construct two environments: a pure simulation environment based on synthetic data and a semi-simulation environment utilizing real-world data.

\subsubsection{Pure Simulation Environment}

In the pure simulation environment, we randomly generate each sampled advertiser's budget $B$ from $U[100, 200]$ for all of her bidding episodes. For each episode, the values $\bmv = (v_1, v_2, \dots, v_n)$ and prices $\bmp = (p_1, p_2, \dots, p_n)$ of the arriving impressions are randomly sampled from predefined distributions. These distributions are conditioned on the generated representation of the advertiser and the index of the impression, ensuring that the cost-performance ratio of the impressions varies over time. Afterward, the impressions are divided into $m$ stages, where the unnormalized proportion of each stage length is sampled from a log-normal distribution with a mean of $e$ and a standard deviation of $0.5$.

To construct the value and price distributions, we generate each sampled advertiser's context $\mathbf{c} = (c_1, c_2) \sim U[0, 1]^2$. For each impression $i \in [n]$, the price $p_i$ is sampled from a log-normal distribution with a mean of $0.1$ and a standard deviation of $0.1$, while the cost-performance ratio $\frac{v_i}{p_i}$ is sampled from a Pareto distribution with a shape parameter of $3 + c_2 \times \cos{\left(2\pi\frac{i}{n} - c_1\right)}$. This ensures that impressions in the middle of the episode have a higher cost-performance ratio than those at the beginning and end. 

Using this environment, we train all the budget planners with $10^6$ randomly generated advertisers and evaluate them on $10^3$ unseen advertisers. We set the number of episodes for each sampled advertiser as $8$ and the number of impressions $n$ as $6000$ for each episode. For all budget planners, we set the number of stages $m = 6$ so that the average cost-performance ratio of impressions in stages $3$, $4$, and $5$ is higher than that in stages $1$, $2$, and $6$.

We implemented two different underlying auto-bidders in the pure simulation environment:
\begin{itemize}
    \item The PID auto-bidder~\citep{bennett1993development}, a classical control-based method that dynamically adjusts the bid amount based on feedback from previous bids and outcomes, aiming to spend the budget smoothly.
    \item The USCB auto-bidder~\citep{he2021unified}, a reinforcement learning-based method that optimizes the bidding strategy under budget constraints. We pretrained the USCB auto-bidder in the pure simulation environment to ensure its effectiveness.
\end{itemize}
For both auto-bidders, the initial budget allocation plan for each advertiser is set according to their budget consumption in all stages of the first episode. As a result, ABPlanner makes its first adjustment at the beginning of episode $1$.

\subsubsection{Semi-Simulation Environment}
\begin{figure*}[t]
    \centering
    \includegraphics[width=0.32\linewidth]{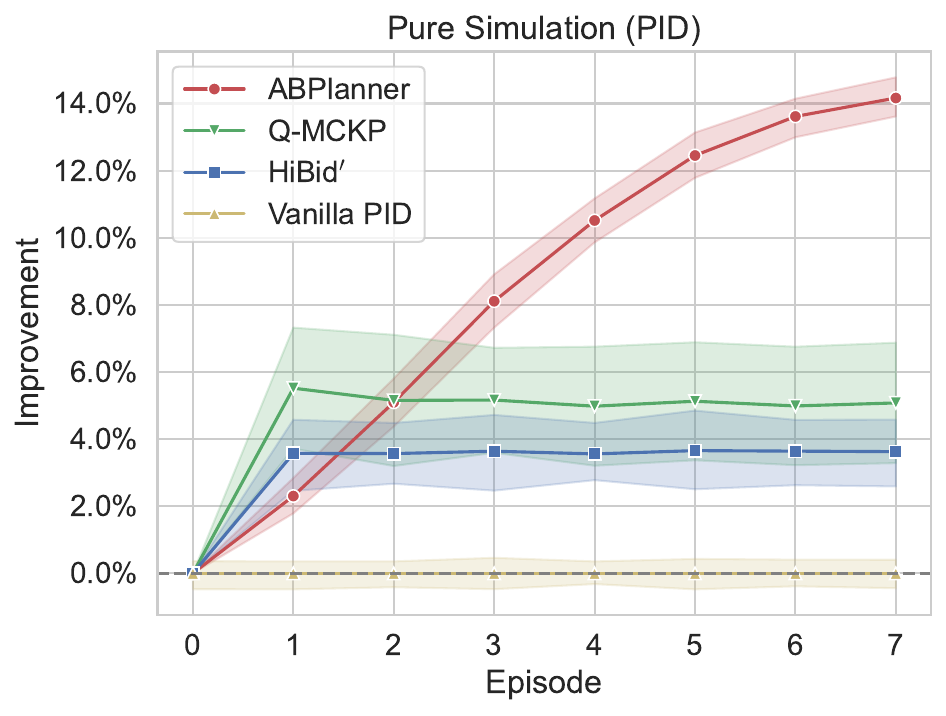}
    \includegraphics[width=0.32\linewidth]{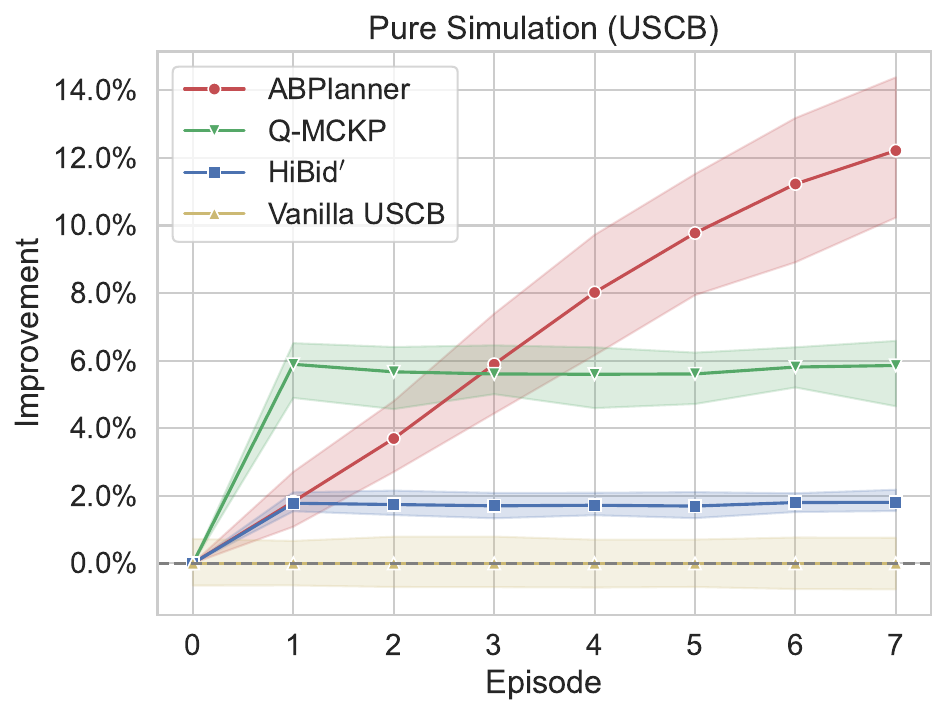}
    \includegraphics[width=0.32\linewidth]{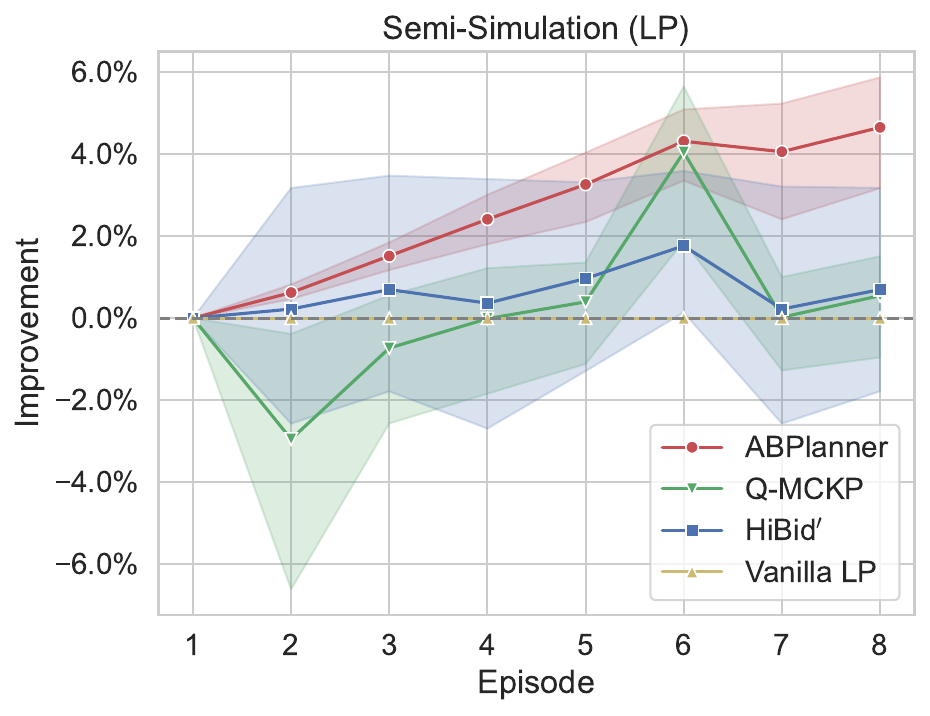}
    \caption{
        Experimental results of simulation experiments.
        We present the return improvement of all the budget planners with respect to vanilla auto-bidders.    
        The average results from $5$ different runs and the $95\%$ confidence intervals are displayed.
    }\label{fig:simulation:return}
\end{figure*}
The semi-simulation environment is built upon bidding logs from a real-world online advertising platform. Initially, we collect the bidding logs of $2 \times 10^4$ advertisers over successive days, setting the predicted conversion rate of an impression as the value of winning it. For each sampled advertiser, we randomly generate her budget $B$ for each bidding episode, treating her bidding logs within one day as a single episode. We randomly divide $90\%$ of the advertisers into the training set and the remaining $10\%$ into the testing set. For all budget planners, we divide the impressions within an episode into $m = 24$ stages, each representing an hour of the day.

In the semi-simulation environment, we implement the underlying auto-bidder as a linear programming-based (LP) auto-bidder. This LP auto-bidder bids based on the hindsight optimal bidding strategy of the previous episode. Formally, given episode $t > 1$ and stage $i \in [m]$, the bid for impression $j$ is $b^t_{i,j} = \lambda^{t-1}_i v^t_{ij}$, where $\lambda^{t-1}_i$ is the optimal bid scaling factor of stage $i$ in episode $t-1$. We solve for $\lambda^{t-1}_i$ using the dual linear programming problem~\citep{he2021unified} of \cref{eq:OPT:bidding}, with values $\bmv^{t-1}_i$ and prices $\bmp^{t-1}_i$ from the previous episode and budget $\rho^t_i$. For the initial episode, we allocate the budget evenly across all stages. Consequently, the LP auto-bidder starts at episode $1$, and all budget planners make their first adjustment at the beginning of episode $2$.

\subsubsection{Baselines} 
To evaluate the effectiveness of ABPlanner in simulation experiments, we compare it with the following budget planning algorithms:
\begin{itemize}
    \item Q-MCKP~\citep{li2018efficient}, which discretizes the budget space into several bins and learns the Q-value function of all stages independently. Based on the learned Q-value, it solves a multi-choice knapsack problem to determine the budget allocation plan.
    \item HiBid$^\prime$, a simplified version of HiBid~\citep{wang2023hibid}, which uses a hierarchical reinforcement learning algorithm to learn both high-level budget planning and low-level bidding strategies. Since we focus on enhancing a fixed underlying auto-bidder, we implement the simplified version HiBid$^\prime$ that only learns the high-level budget planner. This planner, based on reinforcement learning, determines the allocated budget for each stage of a bidding episode.
\end{itemize}
Compared to Q-MCKP, ABPlanner computes the budget allocation plan using neural networks, without the need to discretize the budget space and learn the Q-value function for each stage, which is difficult to learn as discussed in \cref{sec:AdaptableBudgetPlanner}. Moreover, compared to HiBid$^\prime$, which allocates the budget for each episode independently, ABPlanner uses the few-shot interaction data to improve the budget allocation plan over consecutive episodes. 

In addition to these baselines, we also compare ABPlanner with the underlying auto-bidder without any budget planner (which we call the vanilla auto-bidder).

\subsubsection{Implementation}
Our training regimen involves training the high-level budget planner on sampled advertisers and evaluating its performance on unseen advertisers. For a sampled advertiser $c$, we multiply her total budget $B$, budget allocation plan $\bm{\rho}$, cumulative returns $\bm{R}_c$, and cumulative costs $\bm{C}_c$ by $m/B$ when they are used as inputs for all the budget planners.

For both the policy and networks of ABPlanner, we use the same historical embedding $\bmh_t$ in \cref{eq:GRU}. We set $\MLP_e$ as a one-layer neural network with a $64$-dimensional output and ReLU activation function, and we employ a one-layer GRU with a $128$-dimensional hidden state. Based on the GRU output, we use two separate two-layer MLPs with $64$ hidden units and ReLU activation functions to compute the policy and value functions, respectively. The learning rate is fixed at $3 \times 10^{-4}$, while the PPO-clip algorithm's clip parameter $\epsilon$ remains at $0.2$.
For Q-MCKP, following the original paper, we set the number of bins $n_b$ to $40$. For each stage $i \in [m]$ in episode $t > 0$, we set the state $s_t = (\bm{\rho}_i^{0:t-1}, \bm{R}_{c,i}^{0:t-1}, \bm{C}_{c,i}^{0:t-1})$ and fit the Q-value function $Q_i(s_t, \rho_j)$ for each discrete budget $\rho_j$. The architecture of the Q-value network, similar to the value network of ABPlanner, includes a GRU to encode the state and an MLP to estimate the Q-value function.
For HiBid$^\prime$, when determining the budget for stage $i \in [m]$, we set the planner's state as the stage number, the consumed budget, and the remaining budget. We use a PPO algorithm to optimize the planner.

\subsubsection{Experimental Results}
The experimental results of the simulation experiments are presented in \cref{fig:simulation:return}, where we demonstrate the improvement of all budget planners over the underlying auto-bidders.

In the pure simulation experiments (\cref{fig:simulation:return}, left and middle), all budget planners consistently outperform the vanilla PID and USCB auto-bidders, demonstrating the effectiveness of the hierarchical bidding framework. However, the cumulative returns of the baseline planners quickly plateau and fail to show consistent improvement. In contrast, ABPlanner consistently enhances performance across successive episodes and achieves the highest return after episode $3$. These results illustrate the adaptability advantage of ABPlanner, which improves budget allocation through limited interaction with the environment.

In the semi-simulation experiments (\cref{fig:simulation:return}, right), the environment exhibits greater volatility compared to the pure simulation environment, with all approaches experiencing fluctuations across episodes. This increased volatility is likely due to the higher uncertainty of impressions, as they are derived from real-world bidding logs. Despite this volatility, both ABPlanner and HiBid$^\prime$ consistently outperform the vanilla LP auto-bidder. ABPlanner demonstrates more stable performance than the baseline planners and achieves the highest return in all episodes.

\subsubsection{Analysis of Budget Allocation Plan}

To demonstrate how ABPlanner enhances the underlying auto-bidder, we present the average budget allocation plan output by ABPlanner. The averages are calculated across all advertisers in the test set. 

\begin{figure}[t]
    \centering
    \includegraphics[width=\linewidth]{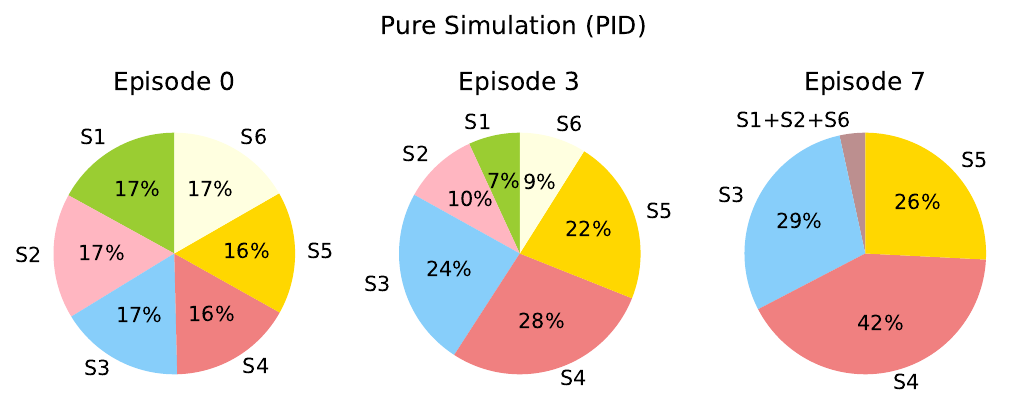}
    \includegraphics[width=\linewidth]{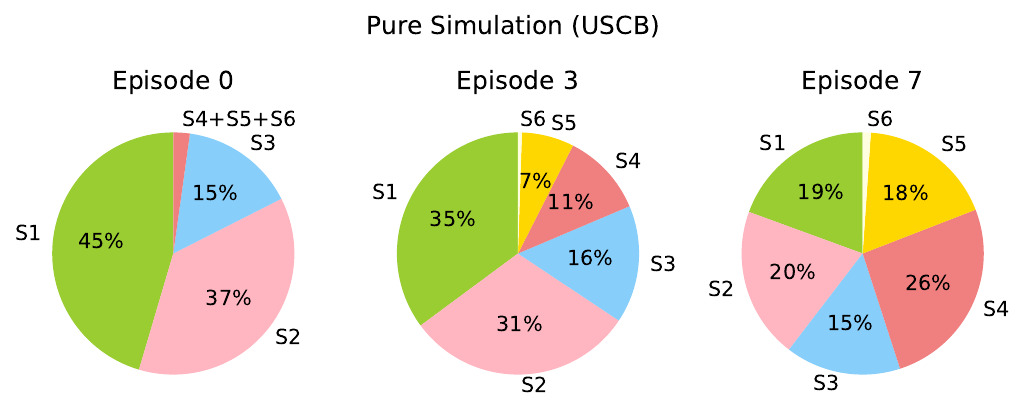}
    \caption{
        The average budget allocation plans output by ABPlanner across all advertisers in the test set of the pure simulation environment.
    }
    \label{fig:pure:budget}
\end{figure}
\cref{fig:pure:budget} presents the results from the pure simulation environment. As illustrated, the vanilla PID auto-bidder allocates a relatively equal budget across all stages, while the USCB auto-bidder allocates more budget in the earlier stages. As the episode progresses, the proportion of the budget allocated to stages $1$, $2$, and $6$ gradually decreases, while the proportion allocated to stages $3$, $4$, and $5$ increases. This strategy aligns with our expectations, as we intentionally set the average cost-performance ratio of impressions higher in stages $3$, $4$, and $5$ than in stages $1$, $2$, and $6$. Thus, ABPlanner effectively captures the temporal variation in cost-performance ratios and strategically allocates more budget to stages with higher expected ratios.

\begin{figure}[t]
    \centering
    \includegraphics[width=\linewidth]{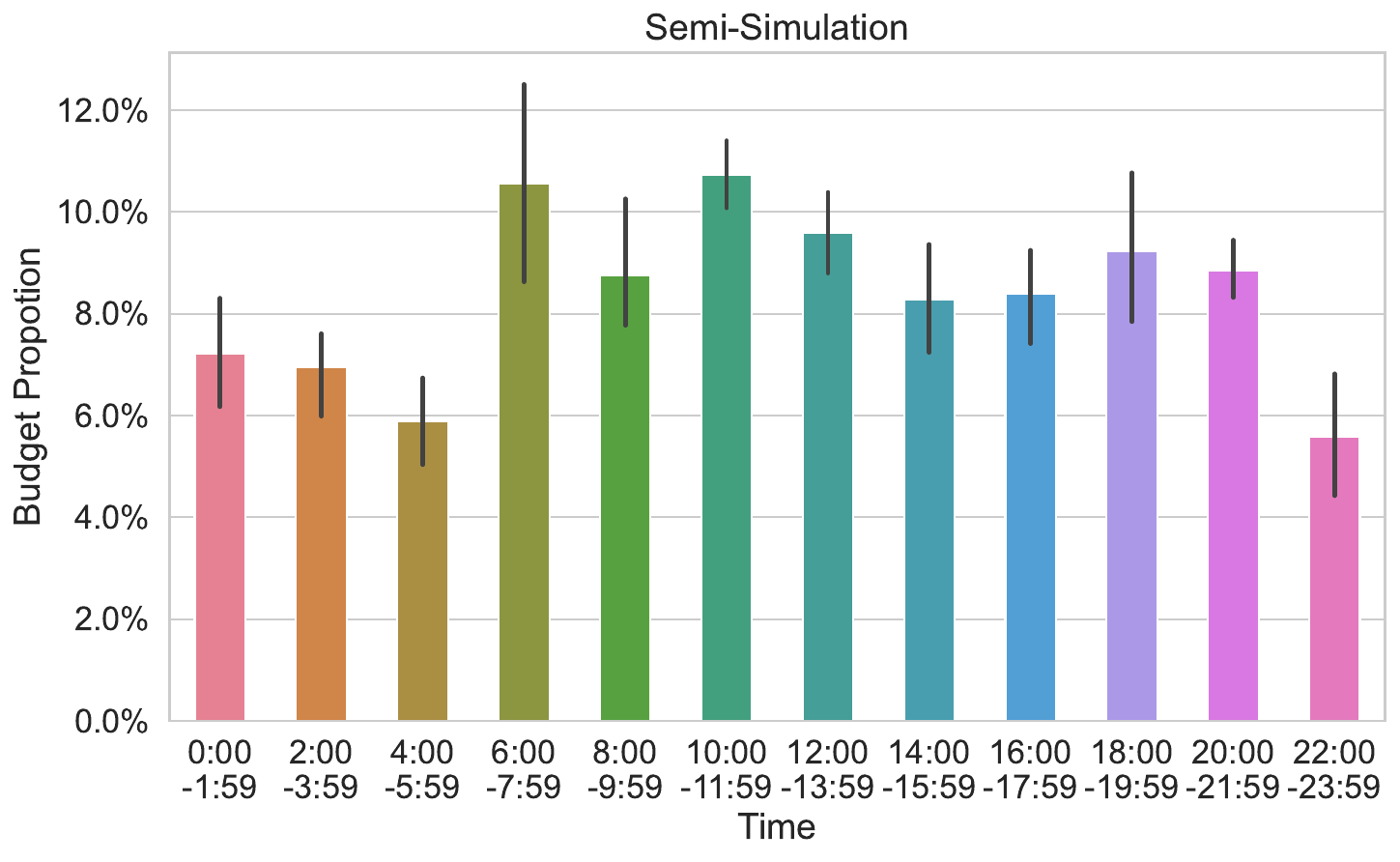}
    \caption{
        The average budget proportion in the last episode of the semi-simulation environment. The proportion is averaged from all advertisers in the test set. Each $2$ successive stages are grouped, displaying the budget proportion for all $12$ groups.
    }
    \label{fig:semi:budget}
\end{figure}
\cref{fig:semi:budget} presents the average budget proportion of the last episode of the semi-simulation experiments. We group each $2$ successive stages and present the budget proportion for all the $12$ groups. As depicted, ABPlanner allocates a higher budget to stages between 6:00 and 21:59 while allocating less budget to the remaining stages, especially those between 2:00 and 5:59. This observation aligns with the real-world trend where the cost-performance ratio of impressions tends to be higher during daytime hours compared to nighttime. This finding provides insight into why ABPlanner effectively adapts to unseen advertisers: on average, stages with high cost-performance ratios across advertisers exhibit similar characteristics. Thus, by strategically allocating more budget to stages with higher cost-performance ratios, ABPlanner achieves improvements. This strategic allocation also underscores ABPlanner's capacity to adapt to temporal variations in cost-performance ratios.

\subsubsection{The Impact of the Number of Stages}
\label{sec:stage}

\begin{figure}[t]
    \centering
    \includegraphics[width=\linewidth]{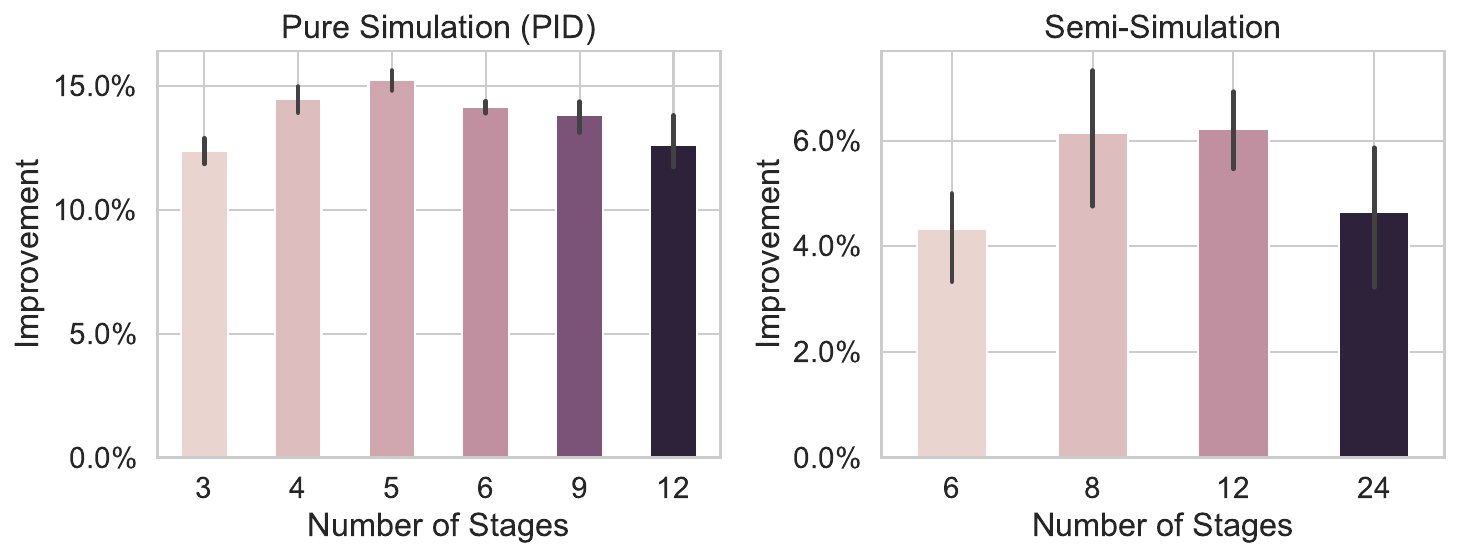}
    \caption{
        Experimental results of the last episode return improvement of ABPlanner with different numbers of stages in the pure simulation and semi-simulation environments.
    }\label{fig:return:stage}
\end{figure}

We investigate the impact of the number of stages $m$ on the performance of ABPlanner. Specifically, we set $m$ to $\{3, 4, 5, 6, 9, 12\}$ in the pure simulation environment with the PID auto-bidder and to $\{6, 8, 12, 24\}$ in the semi-simulation environment. The results are shown in \cref{fig:return:stage}, which displays the improvements in returns achieved by ABPlanner in the last episode compared to the vanilla auto-bidders.

From \cref{fig:return:stage}, it is evident that ABPlanner consistently enhances cumulative returns across all tested values of $m$. Notably, choosing a moderately-sized $m$ yields the best performance (with $m \in \{4, 5, 6\}$ in the pure simulation environment and $m \in \{9, 12\}$ in the semi-simulation environment). This finding indicates that an optimal number of stages strikes a balance between granularity and manageability.

\subsection{Real-World A/B Testing}

We conduct real-world A/B testing within one of the world’s largest e-commerce platforms. This involves comparing the performance of the vanilla auto-bidder with its ABPlanner-enhanced version, setting the number of advertisers to $1000$ and maintaining identical budget allocations across both versions.

\subsubsection{Development Process and Technical Trade-Offs}

To deploy ABPlanner in the online advertising system, we first train it using an offline simulation environment similar to that used in the semi-simulation experiment, where the \emph{predicted conversion rate} is defined as the value of winning an impression. Bidding logs from a single day are treated as a bidding episode, with impressions divided into $24$ stages, each representing an hour of the day.
The vanilla auto-bidder used in the real-world A/B testing resembles the linear programming-based auto-bidder employed in the semi-simulation experiment. However, a key difference lies in the LP auto-bidder's computation of the hindsight optimal bid scaling factor. In the real-world A/B testing, this factor is recalculated at each update step of the auto-bidder (e.g., every $15$ minutes) using bidding logs from the previous day. The budget allocation plan provided by ABPlanner serves as a reference for the budget consumption rate of the underlying auto-bidder.

Real advertising systems have a lower tolerance for performance loss. Therefore, to manage the trade-off between performance and stability, we constrain the magnitude of adjustments made by ABPlanner to prevent significant performance degradation. Additionally, we observe that online impression opportunities exhibit greater variance compared to those in our simulation experiments, especially for advertisers with low budgets. To address this, we adjust ABPlanner to make more conservative budget allocation adjustments for advertisers with lower budgets, considering the variance scale. Moreover, in real-world advertising systems, it is crucial to consider the rationality of allocation schemes based on domain knowledge. Consequently, we implement certain constraints in ABPlanner to prevent extreme or unreasonable allocation scenarios, ensuring a balanced trade-off between optimization and practical feasibility.

\subsubsection{Experimental Results}
\begin{figure}
    \centering
    \includegraphics[width=\linewidth]{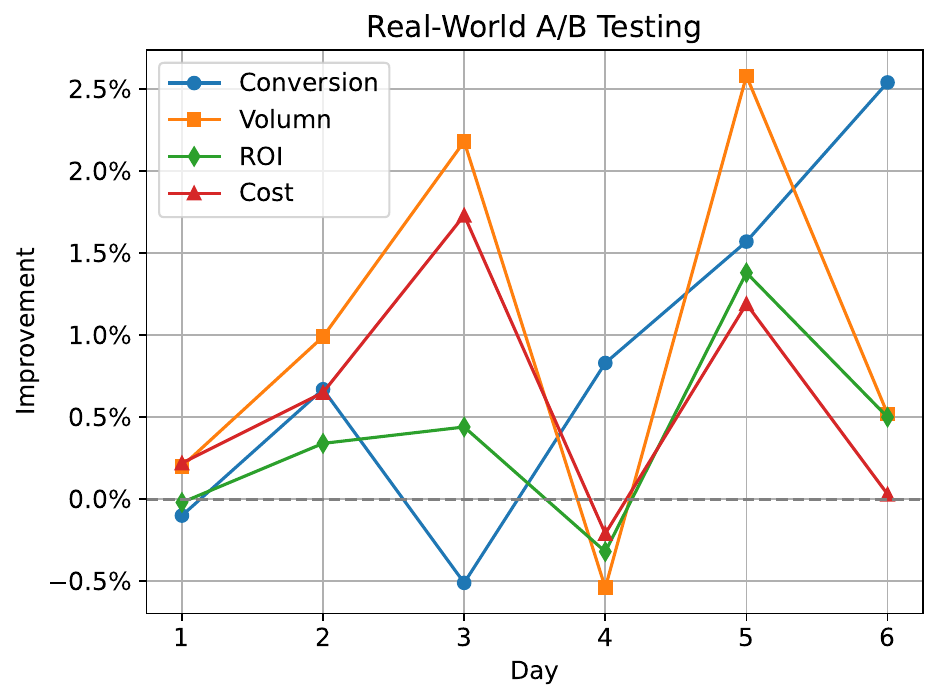}
    \caption{
        Performance improvement achieved by the ABPlanner-enhanced version over the vanilla auto-bidder across six successive days in real-world A/B testing. 
        }
    \label{fig:AB:return}
\end{figure}
\cref{fig:AB:return} illustrates the performance enhancement of the ABPlanner-enhanced version compared to the vanilla auto-bidder over six successive days. 
The metrics examined include the number of conversions, trading volume, return on investment (ROI), and total cost for both versions.

The data indicates that the ABPlanner-enhanced version outperforms the vanilla auto-bidder across most metrics. 
Notably, during the final two days, the ABPlanner-enhanced version demonstrates superior performance in the number of conversions, highlighting the sustained superiority of the ABPlanner-enhanced approach. 
Additionally, ABPlanner improves the total cost on most days, suggesting its capability to indirectly enhance the ad platform's revenue. 
Furthermore, a consistent increase in conversions is observed from the fourth day onwards. 
This observation aligns with the underlying principle of ABPlanner, which is trained using predicted conversion rates as impression values.

In summary, all the experimental results underscore the remarkable adaptability of ABPlanner. 
Even with limited interactions, ABPlanner demonstrates its ability to surpass the performance of the vanilla auto-bidder, showcasing its efficacy in optimizing ad bidding strategies for previously unseen advertisers during testing phases.

\section{Conclusion and Future Work}
\label{sec:Conclusion}

In this paper, we introduce a few-shot adaptable budget planner called ABPlanner to enhance budget-constrained auto-bidding in online advertising. 
ABPlanner is based on a hierarchical bidding framework, which divides the bidding episode into stages and integrates a high-level budget planner to provide budget allocation plans for all stages. 
This approach effectively captures the temporal variation of impressions through the high-level budget planner and mitigates decision-making complexities for the underlying auto-bidder.
Within the hierarchical framework, ABPlanner is modeled as a sequential decision-maker within a Markov Decision Process (MDP), where each bidding episode is treated as one decision step, and each adjustment of the allocation plan is considered an action. 
This design allows ABPlanner to dynamically adjust its budget allocation plan for an advertiser based on the prompts of few-shot episode data collected during interactions.
Through extensive experiments, we demonstrate the effectiveness and adaptability of our approach. 
Furthermore, deployment and validation within the real-world advertising system provide practical confirmation of the utility of ABPlanner.

For future work, since the primary focus of our paper is to develop a high-level budget planner to enhance the underlying auto-bidder without constraining the auto-bidder's specific bidding strategy, an exciting avenue for future research lies in exploring the joint learning of an adaptable budget planner and the underlying auto-bidder. 
This approach holds promise for achieving more integrated and optimized bidding strategies in online advertising.
Additionally, while ABPlanner currently determines the budget allocation plan for all stages in advance before the bidding episode, an intriguing direction for future investigation is to design an adaptable budget planner that dynamically adjusts the budget allocation plan for each stage in sequence. 
This would enable a more fine-grained and adaptive approach to budget allocation, improving the efficiency and effectiveness of online advertising campaigns.

	\section*{Acknowledgments}
		This work is supported by the National Natural Science Foundation of China (Grant No. 62172012), supported by Alibaba Group through Alibaba Innovative Research Program, and supported by Peking University Alimama Joint Laboratory of AI Innovation. 
		We thank all anonymous reviewers for their helpful feedback.
	
	\balance
	\bibliographystyle{plainnat}
	\bibliography{paper/reference}
	
\end{document}